# Quantum properties of dichroic silicon vacancies in silicon carbide

Roland Nagy,[1,†] Matthias Widmann,[1,†] Matthias Niethammer,[1] Durga B.R. Dasari,[1] Ilja Gerhardt,[1,2] Öney O. Soykal,[3] Marina Radulaski,[4] Takeshi Ohshima,[5] Jelena Vučković,[4] Nguyen Tien Son,[6] Ivan G. Ivanov,[6] Sophia E. Economou,[7] Cristian Bonato,[8] Sang-Yun Lee,[9]* and Jörg Wrachtrup[1,2]

[1]*3rd Institute of Physics, University of Stuttgart and Center for Integrated Quantum Science and Technology, IQST, Pfaffenwaldring 57, D-70569 Stuttgart, Germany*

[2]*Max Planck Institute for Solid State Research, Heisenbergstrasse 1, D-70569 Stuttgart, Germany*

[3]*Naval Research Laboratory, Washington, D.C. 20375, USA*

[4]*E. L. Ginzton Laboratory, Stanford University, Stanford, California 94305, USA*

[5]*National Institutes for Quantum and Radiological Science and Technology, Takasaki, Gunma 370-1292, Japan*

[6]*Department of Physics, Chemistry, and Biology, Linköping University, SE-58183 Linköping, Sweden*

[7]*Department of Physics, Virginia Polytechnic Institute & State University, Blacksburg, VA 24061, USA*

[8]*Institute of Photonics and Quantum Sciences, SUPA, Heriot-Watt University, Edinburgh EH14 4AS, United Kingdom*

[9]*Center for Quantum Information, Korea Institute of Science and Technology, Seoul, 02792, Republic of Korea*

Although various defect centers have displayed promise as either quantum sensors, single photon emitters or light-matter interfaces, the search for an ideal defect with multi-functional ability remains open. In this spirit, we study the dichroic silicon vacancies in silicon carbide that feature two well-distinguishable zero-phonon lines and analyze the quantum properties in their optical emission and spin control. We demonstrate that this center combines 40% optical emission into the zero-phonon lines showing the contrasting difference in optical properties with varying temperature and polarization, and a 100% increase in the fluorescence intensity upon the spin resonance, and long spin coherence time of their spin-3/2 ground states up to 0.6 ms. These results single out this defect center as a promising system for spin-based quantum technologies.

# I. INTRODUCTION

Quantum technologies based on solid-state devices can take advantage of well-established fabrication and control methods developed over the past century. Among several quantum systems, color centers in diamond [1–3] have gained prominence as quantum-enhanced nanoscale sensors [4], coherent spin-photon/phonon interfaces [5–7] and quantum registers [8]. Despite their success, the limited emission rate of indistinguishable photons of e.g. the nitrogen vacancy (NV) center and the difficulties of diamond nano-fabrication currently inhibit the progress towards efficient and scalable spin-photon interfacing devices [9] which is a prerequisite for building quantum networks and network-based quantum computing devices. Defect spins in silicon carbide (SiC) have been studied as an analog to diamond color centers, due to their promising complementary properties and the established technologies in growth, doping and device fabrication [10]. As in diamond, defect spins in SiC exhibit long coherence times [11–13] and optically detectable spin signals at room temperatures [14–16], down to the individual spin level [17,18]. SiC hosts several defects with addressable electronic spins, including silicon vacancies [13,18], divacancies [12,17], and transition metal impurities [19,20].

# II. DICHROIC SILICON VACANCY

The silicon vacancy ($V_{Si}$) in SiC is one of the naturally occurring point defects [21] and can be created by kicking out silicon atoms using accelerated particles [14]. There have been two competing models for their atomic structure, an isolated negatively charged $V_{Si}$ [21,22] and a $V_{Si}$ bonded to a neutral carbon vacancy [16,23], for the precise identity of $V_{Si}$. While most defects in semiconductors used in quantum technology host a $S=1/2$ or 1 electronic spins [3,20,22,24], the silicon vacancy ($V_{Si}$) in hexagonal SiC features a $S=3/2$ electronic spin in uniaxial crystal lattices. Its ground state was assigned as S=1 [14] but identified as S=3/2 by many experimental evidences [23,25–27]. According to Kramers' theorem [28], the degeneracy of a half-integer spin system can only be broken by magnetic fields, making it insensitive to fluctuations in strain, temperature, and electric field. Furthermore, the same Landé *g*-factor of ground and excited states makes the optical transitions corresponding to different spin states spectrally indistinguishable, for any applied magnetic field [21]. These factors have led to a theoretical proposal, by Soykal *et al.* [29,30], of a robust interface between spin and photon polarization, which is not perturbed by environmental noise. Additionally, while other defects exist in several different orientations in the crystal lattice [31,32], $V_{Si}$ at each inequivalent lattice exhibit only one single spin orientation along the c-axis of the crystal [14,16]. This can allow deterministic orientation, enhancing scalability in devices. In this work, we demonstrate that the V1 center, one of the two $V_{Si}$ centers residing at two inequivalent lattice sites of in 4H-SiC [21], features a large fraction of radiation into the zero-phonon-lines (ZPL) of up to 40%. In addition, two sharp ZPLs exhibit contrasting polarization properties which may provide an alternate way for quantum control. To discriminate the V1 center with the V2 center whose ZPL is known to be monochromatic, we call it "dichroic silicon vacancy" through this report. These properties can form the basis of the robust spin-photon interface [29]. We also demonstrate efficient spin polarization and readout resulting in nearly a 100% relative increase in optically detected spin signal allowing the high-fidelity spin state readout, and long spin coherence time into the millisecond range. We conclude with some considerations about the prospects to realize a robust spin-photon interface [29,30]. While the V2 center has been intensively studied in the context of quantum applications [13,18,23,27,33–35], there are only a small number of prior studies for the V1 center mostly due to the absence of spin resonance signals at the elevated temperatures [14,21,33].

## III. MATERIALS AND METHODS

All measurements were performed on a commercially available high purity 4H-SiC substrate. The sample was electron irradiated (2 MeV) with a dose of $5\times10^{17}$ electrons/cm$^2$ to create a high density of $V_{Si}$ defect centers. The sample was placed in a closed-cycle cryostat from Montana instruments, at a temperature around 5K. A static magnetic field $B_0 = 60$ and $B_0 = 1000$ G was applied parallel to the c-axis by a permanent magnet. Optical excitation was performed either resonantly, by a 858/861 nm laser diode using a Littrow external cavity or off-resonantly by a 730nm laser diode. The light was focused on the sample by a high NA (0.9) air objective. The RF/MW fields were created by a vector signal generator (Rhode & Schwartz SMIQ 06B), amplified by a 30 W amplifier (Mini-Circuits, LZY – 22+). Radio-frequency (RF) fields were delivered by a copper wire with a diameter of 20 µm, which was spanned on the sample surface. Further information can be found in the Supplemental Material [36].

# IV. FLUORESCENCE PROPERTIES

The theoretical energy-level scheme, proposed by the group-theoretical analysis, is sketched in Fig. 1(a). The ground state of V1 is a spin quartet of symmetry $^4A_2$ ($a_1^2 a_1^1 e^2$) and total spin S=3/2 [21]. The ground $|\pm 1/2\rangle$ and $|\pm 3/2\rangle$ sublevels of V1 are split by a zero-field splitting (ZFS) of 4 MHz [21]. Two excited states $^4E$ ($a_1^1 a_1^1 e^3$) and $^4A_2$ ($a_1^2 a_1^1 e^2$) can be selectively excited from the ground state via resonant laser excitation with 1.445 eV (858 nm) and 1.440 eV (861 nm), known as V1' and V1 ZPLs, respectively [21,37]. At a temperature of 5.5 K, both the emission of the V1 ZPL transition ($^4A_2$ to $^4A_2$) and V1' ZPL emission ($^4E$ to $^4A_2$) are observable as shown in Fig. 1(b) and their decay times, the excited state lifetimes, are approximately 6 ns [36]. Their intensities show temperature dependence, i.e. V1' ZPL intensity peaks at around 70 K (Fig. 1(b)). The energy difference between the ZPLs of V1 and V1' is about 4.4 meV which corresponds to a thermodynamic equivalent temperature of 51 K. The enhanced emission from the V1' transition at elevated temperatures may be understood as a phonon-assisted process [36].

The protocol by Soykal *et al.* for a robust spin-photon interface features the energetically degenerate but orthogonally polarized photons [29,30] associated to V1. Here we report a complete characterization of the polarization properties of V1' and V1 ZPL. While some polarization studies are reported in the literature [21,38], we will show that the current model for this defect requires revision. The polar plot in Fig. 1(c) represents the integrated intensity of each of the V1 and V1' ZPLs as a function of the half-wave plate angle, taken at T=5.5 K with the laser incident angle perpendicular to the c-axis. The dominant polarizations of V1 and V1' ZPLs are almost orthogonal to each other. The full orientation analysis results are in qualitative agreement with previously suggested optical selection rules based on group-theoretical analysis within the single group $C_{3v}$, representing the symmetry of $V_{Si}$ [21,38]. This analysis predicts $\mathbf{E} \parallel \mathbf{c}$ polarization for the V1 transition and $\mathbf{E} \perp \mathbf{c}$ polarization for the V1' transition. However, while the photons originating from V1' are quite well linearly polarized $\mathbf{E} \perp \mathbf{c}$, the V1 transition is not entirely polarized as $\mathbf{E} \parallel \mathbf{c}$ but contains a component $\mathbf{E} \perp \mathbf{c}$. These indicate that the selection rules need revision. Since the negatively-charged $V_{Si}$ contains an odd number of electrons (resulting in half-integer spin), the correct symmetry is the double group $\overline{C_{3v}}$, as previously suggested [29]. The derivation of the selection rules for $\overline{C_{3v}}$ leads to a better estimate of the relative contribution of the $\mathbf{E} \parallel \mathbf{c}$ and $\mathbf{E} \perp \mathbf{c}$ polarizations in the optical emission of the V1 and V1' ZPLs. This is outlined in the Supplemental Material [36]. We find that for the V1 transition the distribution among the two polarizations is $\mathbf{E} \parallel \mathbf{c} : \mathbf{E} \perp \mathbf{c} = 3:1$, whereas for the V1' transition the proportion is $\mathbf{E} \perp \mathbf{c} : \mathbf{E} \parallel \mathbf{c} = 11:1$. These estimates are in good agreement with the polar plots in Fig. 1(c), $\mathbf{E} \parallel \mathbf{c} / \mathbf{E} \perp \mathbf{c} = 1.85 \pm 0.06$ for V1, and $\mathbf{E} \perp \mathbf{c} / \mathbf{E} \parallel \mathbf{c} = 19 \pm 3$ for V1'. For completeness, polarization is also measured with the laser incident angle parallel to the c-axis [36].

## V. OPTICAL SPIN STATE DETECTION

In the next set of measurements, we characterize the spin properties for a defect ensemble. The ground state spin can be polarized into the sublevels $|S_z = \pm 3/2\rangle$ [29] by optical pumping [33] with an off-resonant laser ($\lambda$ =730 nm). At any constant finite magnetic fields ($B_0$), the spin energy levels are determined by the Hamiltonian,

$$H = g\mu_B \mathbf{B} \cdot \mathbf{S} + D\left[S_z^2 - S(S+1)/3\right], \quad (1)$$

where $g$ is the Landé $g$-factor, $\mu_B$ is the Bohr magneton, $D$ is the ZFS ($2D = 4$ MHz), and $S_z$ is the projection of the total spin onto the quantization axis, the c-axis in 4H-SiC. By applying resonant radio-frequency (RF) fields ($B_1$) one can induce transitions between spin sublevels (Fig. 2(a)), resulting in a change in optically detected magnetic resonance (ODMR) as shown in Fig. 2(b). The relative change in ODMR signal is calculated as $\left[I(f) - I_{off}\right]/I_{off}$, where $I(f)$ is the PL intensity at the RF frequency $f$, and $I_{off}$ is the PL intensity at the off-resonant RF frequency. The spin-sublevels are energetically split at $B_0 = 60$ G aligned along the c-axis. The relative ODMR signal as a function of the driving RF frequency shows a negative signal at 170 MHz, with the relative intensity 0.05% as in the upper panel of Fig. 2(b). It is attributed to the V1 ground state spin; a similar signal was also reported for V1 and V3 centers in 6H-SiC [33]. By exciting the optical transition V1 resonantly a positive relative ODMR signal with $100 \pm 0.6\%$ is achieved (lower panel in Fig. 2(b)). In contrast, excitation of the V1' optical transition in Fig.1(a) reveals a negative signal with a minimal change in the relative signal intensity [36]. A similar substantial enhancement of the ODMR signal was reported for the V2 center ensemble in 6H-SiC [33]. Although the underlying mechanism is not yet completely understood, we attribute it to the enhanced spin polarization in the ground state resulting from resonant optical excitation. Resonant excitation of V1 ZPL results in the excitation into the lowest vibrational level of the V1 excited state. This efficiently suppresses the phonon-assisted spin-mixing between the $^4A_2$ and $^4E$ excited states leading to an improvement in the ODMR signal at sufficiently low temperatures ($k_BT < 4.4$ meV). On the other hand, resonant excitation of V1' does not result in such an improvement as it still involves the excitation of V1 (lower in energy) and its vibrational levels. This observation may also indicate that optical polarization is mainly established by the intersystem-crossing (ISC) between the $^4A_2$ excited states and the $^2E$ metastable state while the $^4E$ excited states have a less efficient ISC [29]. We expect further enhancement of the ODMR signal when a single V1 center is isolated owing to the suppressed inhomogeneous broadening. We note that if an identical ODMR contrast, namely $C$, and an identical photoluminescence (PL) intensity without spin resonance are assumed, a positive signal leads to a signal-to-noise ratio larger than the negative signal, as of the NV center in diamond [1] and divacancies in 4H- and 6H-SiC [32], by a factor of $(1-C)^{-0.5}$ [36].

## VI. COHERENT SPIN CONTROL

In order to demonstrate coherent control of the electronic spins, we investigated spin dynamics of the V1 center ensemble under applied RF pulses. The resulting distribution of the Rabi oscillations in the range of $f$ =160-190 MHz at $B_0 = 60$ G, is shown in Fig. 3(a). The dynamics is strikingly different from the single-frequency Rabi oscillations typical of a two-level system. Further understanding can be obtained by plotting the Fast-Fourier transform of the Rabi oscillations, for different values of the driving power. For two-level transitions one expects parabolic profiles, corresponding to a Rabi frequency $\Omega$ increasing with the detuning $\Delta\omega$ as $\Omega^2 = \Omega_0^2 + \Delta\omega^2$ where $\Omega_0$ is the driving frequency determined by the applied $B_1$ field strength ($\Omega_0 \propto B_1$). The experimental data reveal richer and more complex dynamics. We explain our observations with the presence of three closely-spaced transitions, corresponding to $f_1$, $f_2$, and $f_3$ in Fig. 2(a). While resonantly driving one transition, due to the small ZFS, off-resonant excitation of the other two transitions is not negligible. To support this explanation, we developed a theoretical model based on four levels of S=3/2 driven by a single monochromatic radio-frequency field. The system dynamics is investigated assuming initial polarization into an incoherent mixture of $|S_z = \pm 3/2\rangle$. Further details on the model can be found in the Supplemental Material [36]. Our simulations match quite closely the complex structure of the experimental data (see Figure 3(c, d)). When $f$ is lower than $f_1$ ($|-3/2\rangle \leftrightarrow |-1/2\rangle$), the transition $f_1$ is mainly excited, leading to a parabolic profile. However, off-resonant excitation of the transition $f_2$, coupling $|-1/2\rangle$ to $|+1/2\rangle$, results in a second weaker Fourier component in the Rabi spectrum with larger Rabi frequency. With increasing RF power ($B_1$ field strength), simulated by increasing the driving frequency proportional to the increase of the experimentally used $B_1$ field strength, this component becomes stronger. When $f=f_2$, one simultaneously drives off-resonantly the transitions $f_1$ and $f_3$, resulting in a larger Rabi frequency. This is evident in the plots corresponding to the largest RF power, where the parabolic profile centered around $f_2$ shows a much larger Rabi frequency than the profiles related to $f_1$ and $f_3$. Note that the experimental data can only be explained by assuming the excitation of the $|-1/2\rangle \leftrightarrow |+1/2\rangle$ transition, which was not reported. Additionally, the assumption for initial polarization into $|\pm 1/2\rangle$, which is the case for the V2 center, does not reproduce the observed signal. Note that we report for the first time the experimental evidence for S=3/2 of the ground state of the V1 in 4H-SiC [36].

The small ZFS poses challenges for high-fidelity coherent spin control, which need to be addressed for the V1 center to be a serious contender for quantum technology. There are several possibilities to explore: use of (i) optimal quantum control sequences, (ii) adiabatic passage techniques that restrict the dynamics only to a two-level subspace (e.g. $|+3/2\rangle$ and $|-1/2\rangle$), with no leakage to other states of the Hilbert space [39], (iii) pulses designed to avoid a transition by building holes in their frequency spectrum to avoid leakage [40], (iv) superadiabatic (shortcuts to adiabaticity) control [41], which was recently demonstrated for NV centers in diamond [42]. Alternatively, the V1 in 6H-SiC is known to exhibit a larger ZFS [14,21], which would relax this problem.

## VII. SPIN DECOHERENCE

We studied spin coherence at T= 5.5 K with $B_0 = 60$ G [36] and 1000 G by Ramsey, Hahn-echo, and XY-8 dynamical decoupling pulse sequences by optical excitation resonant with the V1 ZPL (861 nm). We observed an evolution of the coherent superposition with the electron spin dephasing time of $T_2^* = 1.3 \pm 0.3$ μs at $B_0 = 1000\,\text{G}$ aligned parallel to the c-axis by a Ramsey experiment as shown in Fig. 4(a). To suppress the inhomogeneous broadening in an ensemble and decouple the spin ensembles from low-frequency spin noise from such as paramagnetic impurities and a nuclear spin bath composed of $^{29}$Si and $^{13}$C [11], we applied a Hahn-Echo sequence. Identical laser pulses of 2 μs length were applied before and after the MW pulse sequences for the optical spin polarization and projective spin state readout, respectively, and also to avoid dephasing due to the optical excitation [29]. Although the applied RF pulses exhibit limited spin control to a single transition as discussed in the Supplemental Material [36], we could see a typical exponential decay with $T_2 = 83.9 \pm 1.6$ μs (Fig. 4(b)). The observed $T_2$ is, however, shorter than the theoretical expectations for the V2 center [11] and the value measured for a single V2 center at room temperature [18]. This could be related to the imperfect π pulses and the inhomogeneity of the $B_0$ field (see the Supplemental Material [36]). These observations, however, support the findings by Carter *et al.* [34], related to the fact that the dephased state cannot be refocused by a π pulse due to the oscillating local fields produced by coupled nuclear spins. Thus, the shorter $T_2$ could be related to electron spin echo envelope modulation (ESEEM). The four sublevels of a S=3/2 electronic spin have four different non-zero hyperfine coupling to nearby nuclear spins and thus result in more complex ESEEM than S=1 systems [11,18]. Furthermore, as reported by Carter *et al.*, the ensemble inhomogeneous broadening induces beating oscillations among the various modulation frequencies, leading to a shortening of the Hahn-echo $T_2$ [34]. To further suppress decoherence, we applied the XY-8 dynamical decoupling sequence, which acts as a filter for the environmental magnetic noise [43]. This sequence has proven to be effective to extend the coherence time of the S=3/2 spin ensemble of the V2 center from the nuclear spin bath in 4H-SiC [13]. A repeated decoupling pulse scheme leads to a better suppression of noise, increasing the spin decoherence time with N=10 and N=50 repetitions to a value of respectively $T_2 = 286 \pm 7$ µs and $T_2 = 0.60 \pm 0.01$ ms (Fig. 4(b)). These suggest that the heteronuclear spin bath in SiC itself provides a diluted spin bath for not only the V2 center [13,18], and divacancy defects [17], but also the V1 center.

## VIII. ISOLATED SINGLE SILICON VACANCY

Although the spin ensemble-based quantum applications such as quantum memory [44] are valuable, many advanced quantum applications including e.g. the spin-photon interface requires addressing of single defect centers. To test if the single V1 center can be used as an efficient coherent single photon source, e.g. a building-block for the robust spin-photon interface, we isolated single V1 centers in nanopillars fabricated on a 4H-SiC sample [35], as shown in Fig. 5(a). Addressing of a single center is proven by the autocorrelation measurement in a Hanbury Brown and Twiss configuration, with $g^{(2)} < 0.5$ [36]. The saturated count rate of 14 kcps was measured by an air objective of NA=0.9 with a single photon detector inefficient in this wavelength (see the Supplemental Material [36]). The spectrum at $T$=4K shows both V1 and V1' ZPLs as in Fig. 5(b), further proof that they correspond to two different excited states of the same defect. To determine the Debye-Waller factor (DWF), the fraction of radiation into the ZPL of V1 over the whole V1 spectrum, the contribution of V1' to the phonon sideband (PSB) has to be minimized. At 4 K, the intensity of the V1' ZPL is weak, and we suppose that the contribution to the PSB is also negligible in comparison to V1. Then, the conservatively estimated DWF of the V1 is 40±6 %. See the Supplemental Material [36] for the additional data and DWF of the V1'.

## IX. SUMMARY AND OUTLOOK

In summary, optical spectroscopy and polarization measurements confirm the symmetry properties of the V1 center in 4H-SiC, supported by the established double group $\overline{C}_{3v}$ model, substantiating the theoretical model leading to the proposed robust spin-photon interface [29]. A spin-photon interface requires narrow optical transitions [36], weak spectral diffusion and slow spin-flip rates by optical pumping cycles. Recent results on the divacancy in SiC shows that the material quality is sufficiently high to satisfy these requirements [22]. Resonant optical excitation of the V1 ZPL leads to a substantial increase in spin-dependent photoluminescence emission indicating an efficient spin-dependent transition. We also have shown the extension of spin coherence time, through dynamical decoupling sequences, up to 0.6 ms, which will enable long and complex spin manipulation necessary for the spin-photon interface [5,45]. While the leading contenders for defect-based quantum spintronics, such as the NV center in diamond and divacancy in 4H-SiC, suffer from low optical emission into zero-phonon lines (with DWF ~3% and 5-7% [22,46], respectively), the V1 center in 4H-SiC features a significantly higher DWF, up to 40%. The high ZPL emission could guarantee a high event rate for the proposed generation of spin-photon entanglement. The weak overall photon emission rate of the V1/V1' transition may be circumvented by using photonic structures fabricated on SiC, which recently have shown progress towards high-Q cavities and efficient photon collection [35,37,47]. This can further be used to generate strings of entangled photons [6,48]. Further work on this dichroic single defect, which is ondergoing, is necessary to identify individual optical transitions and the associated selection rules which are essential for realizing spin-photon interfaces, e.g. spin-photon entanglement exploiting transitions to the $^4E$ excited state (V1' line) with the high fidelity spin initialization and read-out with the V1 line.

# ACKNOWLDEGEMENT

This work was supported by the ERA.Net RUS Plus Program (DIABASE), the DFG via priority programme 1601, the EU via ERC Grant SMel and Diadems, the Max Planck Society, the Carl Zeiss Stiftung, the Swedish Research Council (VR 2016-04068), the Carl-Trygger Stiftelse för Vetenskaplig Forskning (CTS 15:339), the Knut and Alice Wallenberg Foundation (KAW 2013.0300), the JSPS KAKENHI (A) 17H01056, the National Science Foundation DMR Grant Number 1406028, the U.S. Office of Secretary of Defense Quantum Science and Engineering Program, the COST Action MP1403 "Nanoscale Quantum Optics" funded by COST (European Cooperation in Science and Technology), and by EPSRC (grant EP/P019803/1), the Army Research Office under contract W911NF1310309, and the KIST Open Research Program (2E27231) and institutional program (2E27110). We thank Roman Kolesov, Rainer Stöhr, and Torsten Rendler for fruitful discussions and experimental aid. We also acknowledge motivating discussions with Michel Bockstedte, Adam Gali, Thomas L. Reinecke, and Jingyuan Linda Zhang. We thank Rainer Stöhr for proofreading.

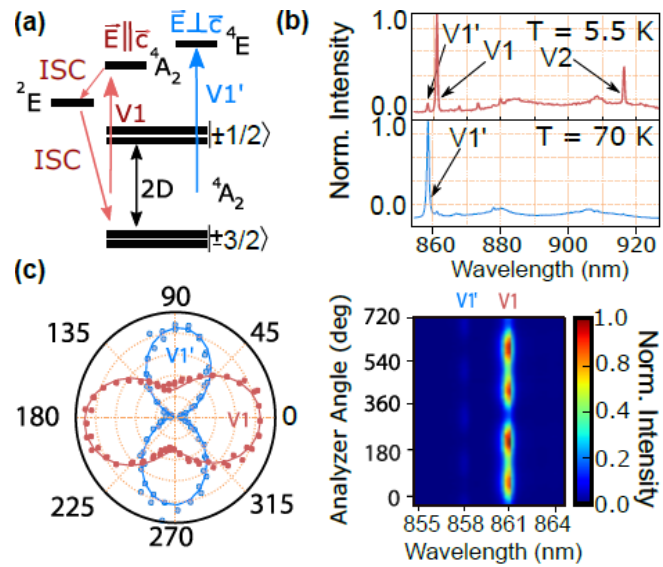

FIG. 1. (a) The energy level scheme of the V1 center. (b) The PL spectrum of a $V_{Si}$ ensemble at 5.5 K and 70 K. (c) The optical polarization of the V1/V1' transitions at the sample orientation in which the c-axis is perpendicular to the laser incident direction at 5.5K. Left: the polar plot of the normalized V1 and V1' intensities. 0°, equivalently 180°, corresponds to the c-axis orientation. Right: the density plot showing the absolute intensities of the V1/V1' ZPLs.

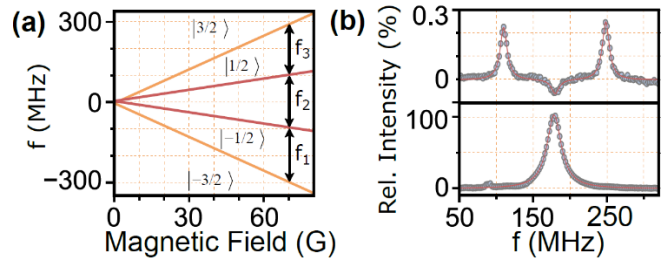

FIG. 2. (a) Zeeman effect of the spin 3/2 ground state of the V1 center for $B_0 \parallel c$. $f1, f2,$ and $f3$ represent possible resonant transitions. (b) Upper panel: an ensemble ODMR spectrum with a 730 nm laser at 60 G. Lower panel: the ODMR with a laser resonant to V1, 861 nm.

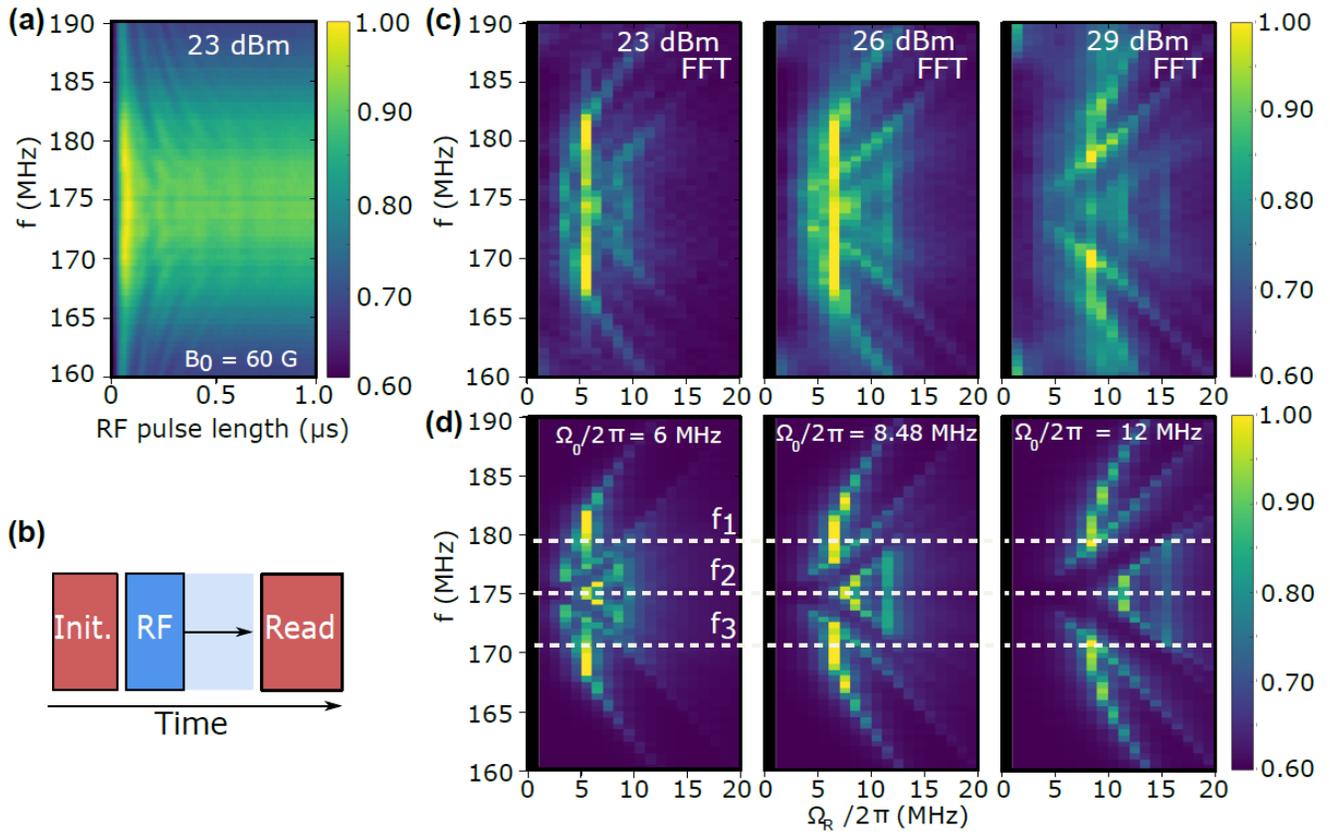

FIG. 3. (a) Rabi measurement with detuned RF driving frequencies. (b) Pulse scheme for a Rabi measurement. The first laser pulse (Init.) is polarizing the spin state. The RF pulse is manipulating the spin state followed by the last laser pulse (Read) for the spin state readout. (c) Fast Fourier transformed Rabi oscillations at different RF powers. (d) Simulated Rabi oscillations. The dotted lines indicate three resonant RF frequencies shown in Fig. 2(a). The strong zero frequency intensities in both (C) and (D) are removed for better distinguishability of the Rabi frequencies.

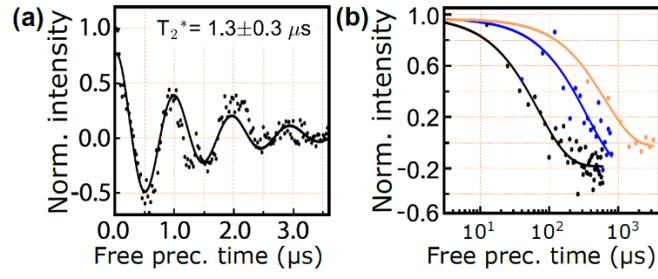

FIG. 4. (a) Ramsey measurement at $B_0 = 1000\,\text{G}$. (b) The spin decoherence measured at $B_0 = 1000\,\text{G}$ by Hahn-Echo (black) and XY-8 dynamic decoupling (blue: N=10, orange: N=50). See the Supplemental Material for the used pulse sequence [36].

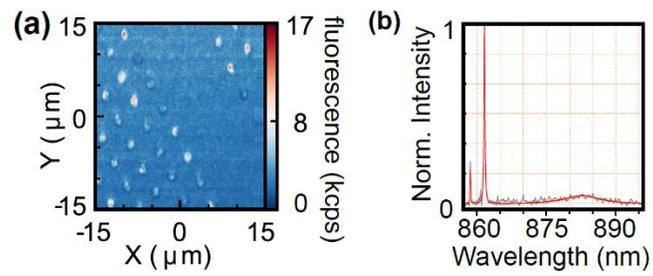

FIG. 5. (a) Confocal fluorescence raster scan showing single silicon vacancy V1 and V2 centers in SiC nanopillars at 4K. (b) A single V1 defect PL spectrum with the V1' and V1 ZPLs at 858 and 861 nm, respectively.

*sangyun.lee@kist.re.kr

†These authors contributed equally to this work.

# Supplemental Material for "Quantum properties of dichroic silicon vacancies in silicon carbide"

Roland Nagy,[1,†] Matthias Widmann,[1,†] Matthias Niethammer,[1] Durga B.R. Dasari,[1] Ilja Gerhardt,[1,2] Öney O. Soykal,[3] Marina Radulaski,[4] Takeshi Ohshima,[5] Jelena Vučković,[4] Nguyen Tien Son,[6] Ivan G. Ivanov,[6] Sophia E. Economou,[7] Cristian Bonato,[8] Sang-Yun Lee,[9]* Jörg Wrachtrup[1,2]

[1]3. Institute of Physics, University of Stuttgart and Center for Integrated Quantum Science and Technology, IQST, Pfaffenwaldring 57, D-70569 Stuttgart, Germany

[2]Max Planck Institute for Solid State Research, Heisenbergstrasse 1, D-70569 Stuttgart, Germany

[3]Naval Research Laboratory, Washington, DC 20375, USA

[4]E. L. Ginzton Laboratory, Stanford University, Stanford, California 94305, USA

[5]National Institutes for Quantum and Radiological Science and Technology, Takasaki, Gunma 370-1292, Japan

[6]Department of Physics, Chemistry, and Biology, Linköping University, SE-58183 Linköping, Sweden

[7]Department of Physics, Virginia Polytechnic Institute & State University, Blacksburg, VA 24061, USA

[8]Institute of Photonics and Quantum Sciences, SUPA, Heriot-Watt University, Edinburgh EH14 4AS, UK

[9]Center for Quantum Information, Korea Institute of Science and Technology, Seoul, 02792, Republic of Korea

## 1. Experimental Method

All measurements were performed on a high purity 4H-SiC substrate purchased from Cree, Inc (2 mm × 1 mm × 0.5 mm). The sample was electron irradiated (2 MeV) with a dose of $5\times10^{17}$ electrons/cm² to create a high density of $V_{Si}$ defect centers. In order to remove contaminations on the surface acetone in an ultrasonic bath was used. All the spin measurements at low magnetic field ($B_0$ = 60 G) were done with the sample flipped on the side, with the c-axis perpendicular to the optical excitation/detection axes.

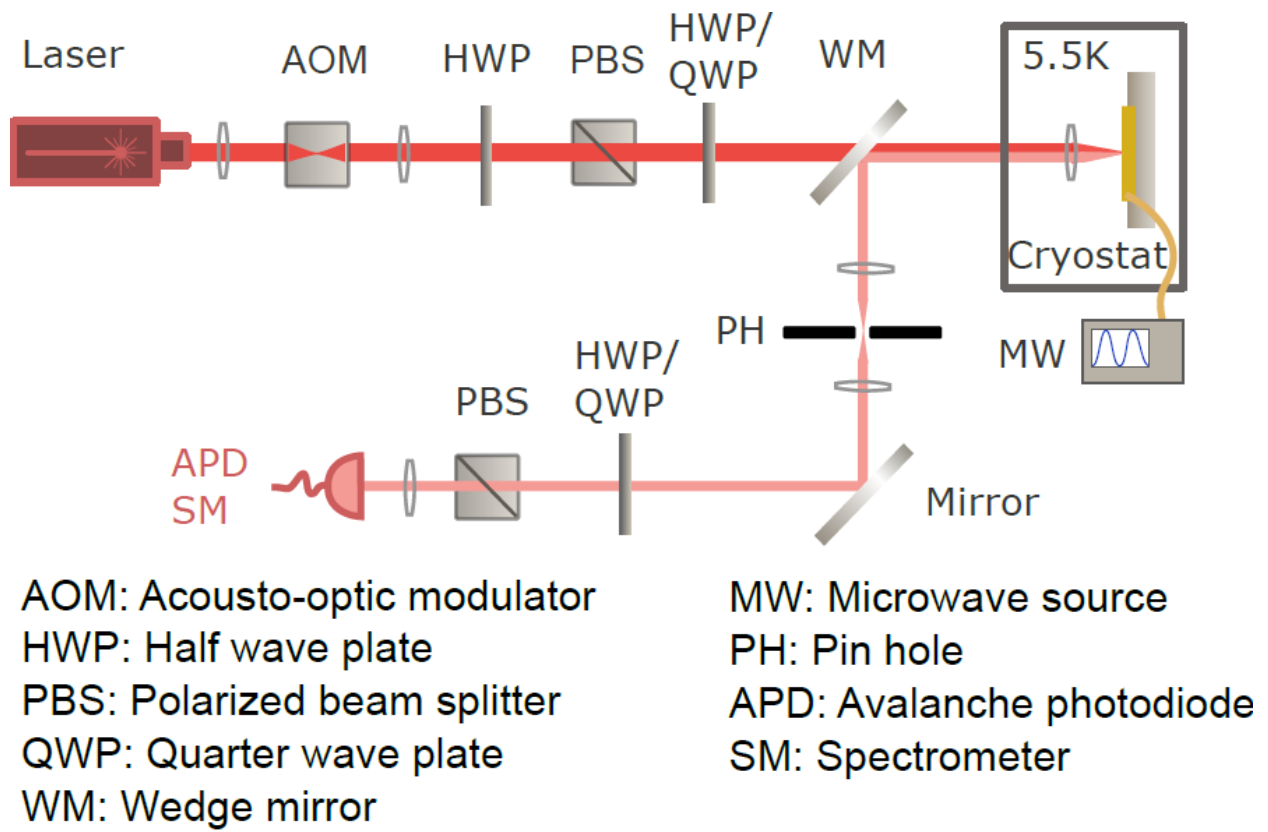

AOM: Acousto-optic modulator
HWP: Half wave plate
PBS: Polarized beam splitter
QWP: Quarter wave plate
WM: Wedge mirror

MW: Microwave source
PH: Pin hole
APD: Avalanche photodiode
SM: Spectrometer

FIG. S1. Experimental setup. The measurements were done with a home-built confocal setup. A closed-cycle cryostat from Montana Instruments was used to perform the low-temperature experiments. See text for details.

The sample was placed in a closed-cycle cryostat from Montana instruments, at a temperature around 5K. For measurements at low magnetic field, a static magnetic field $B_0$=60 G was applied parallel to the c-axis by a permanent magnet outside the cryostat chamber, mounted on an x-y-z-linear translation stage. The magnetic field splits the energy levels of the spin states so that we could address them individually with radio-frequency (RF) electromagnetic wave. For a high magnetic field ($B_0$=1000 G ), a smaller permanent magnet was placed between the sample and the heat sink in the cryostat.

Optical excitation was performed either resonantly, by a 858/861 nm laser diode using a Littrow external cavity (900 µW on the sample, 300 GHz linewidth) or off-resonantly by a 730nm laser diode (usually 500 µW on the sample, or being varied from 0 to 500 µW for the optical saturation as in Fig.6c of the main text). The light was focused on the sample by a high NA (0.9) air objective (Zeiss). Photoluminescence spectra were recorded by a spectrometer (Princeton Instruments, Acton SP2300, grating: 300 g mm$^{-1}$). For the determination of the Debye-Waller factors (Fig. 6 of the main text and the section S7), the changes in the responsivity of the CCD camera and the efficiency of the grating are considered. For both CW and pulsed ODMR experiments using the resonant optical excitation, we sent another re-pumping laser (730 nm), off-resonant to the ZPLs simultaneously into the sample. Re-pumping increased the total PL intensity in the phonon side band slightly by up to a factor of two. However, the ODMR contrast was independent of the use of the additional off-resonant laser.

For polarization and magnetic resonance measurements, the photoluminescence was detected by near infrared enhanced APDs from Perkin Elmer (SPCM-AQRH-15). Photon polarization was analyzed by a half-wave plate (HWP) and a polarizing beam splitter (PBS) in front of the spectrometer. The HWP was rotated by a stepper-motor (M101, LK-instruments) with 0.3° resolution. An 887 nm tunable long pass filter (VersaChrome TLP01-887-25x36) was placed before the detectors.

The RF/MW fields were created by a vector signal generator (Rhode & Schwartz SMIQ 06B), amplified by a 30 W amplifier (Mini-Circuits, LZY – 22+). Radio-frequency (RF) fields were delivered by a copper wire with a diameter of 20 µm, which was spanned on the sample surface. For the creation of RF pulses, an RF switch (ZASWA-2-50DR, Mini-Circuits), controlled by a home-built FPGA-based pulse generator was used. The laser was pulsed by an acousto-optical modulator (EQ Photonic 3200-124) whose driving RF wave was pulsed by an RF switch, also controlled by the home-built pulse generator. The phase for the dynamical decoupling sequence XY-8 was controlled using the vector mode of the SMIQ 06B. The input of the I/Q modulator of the SMIQ 06B was controlled by RF switches from Mini-Circuits and the home-built electronic device for the voltage generation. See Fig. S1 for schematic description.

## 2. Temperature dependence model

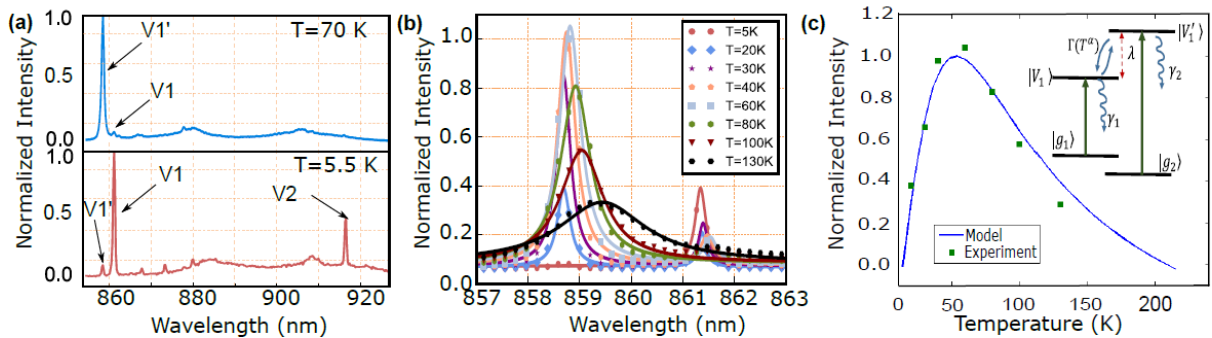

FIG. S2. (a) PL spectra of a VSi ensemble at 5.5 and 70 K. V1' and V1 ZPLs are at 858 nm and 861 nm, respectively, while V2 ZPL is at 916 nm. (b) The temperature dependence of both V1 and V1' ZPLs from 5 K to 130 K. (c) A model to describe the temperature.

In the main text, we presented the PL spectrum of the $V_{Si}$ ensemble at 5.5 K. In this section, we report the temperature dependence of it and discuss a model for the observed temperature dependence of the PL, shown in Fig. S2.

At a temperature of 5.5 K, the emission of the V1 ZPL transition ($^4A_2 |es\rangle$ to $^4A_2|gs\rangle$) dominates. With increasing temperature, the intensity of the V1 ZPL decreases, whereas the V1' ZPL emission ($^4E |es\rangle$ to $^4A_2|gs\rangle$) becomes more prominent and peaks at around 70 K, as shown in Fig. S2. The energy difference between the ZPLs of V1 and V1' is about 4.4 meV which corresponds to a thermodynamic equivalent temperature of 51 K. The enhanced emission from the V1' transition at elevated temperatures may be understood as a phonon-assisted process, where the temperature-induced dephasing of the orbital excited states leads to thermally-induced excitation transfer among the excited states between V1 and V1' excited states. The transfer reaches a maximum when temperature-induced mixing (dephasing) becomes comparable to the splitting between the energy levels. The experimental observation, however, cannot be explained by a simple thermally induced excitation transfer but rather requires a more involved mixing of excited states. We performed numerical simulations of the Lindblad master equation for a four-level system with two ground and two excited states as shown in Fig. S2(c) using the experimental decay rates of the V1 and V1' excited states (to be discussed later). In this model, as discussed below, we introduced, in addition to the normal decay channels from the excited states as shown above, a temperature-dependent mixing of the excited states at a rate $\Gamma(T^\alpha)$ with a power-law dependence on temperature. We find that the best fit to our data occurs for $\alpha \sim 1.57$. This is consistent with the inter-level phonon coupling (pseudo Jahn-Teller effect) between the V1 and V1' states mediated by the E-symmetry vibrational normal modes of the lattice [1].

Starting from the molecular orbits of the defect center, one can form irreducible representation for the symmetry operator of the defect in which the ground and excited states are constructed. By incorporating the spin-orbit and spin-spin interactions, one can find the splitting among various states resulting in the formation of energy subspaces $^4A_2$ ground and excited states separated by 1.44 eV, and the $^4$E excited state split from the $^4$A$_2$ excited state by 4.4 meV. At any finite temperature T, these energy states are broadened; thermal energy acts as a dephasing operation that introduces incoherent mixing among the energy states. As the $^4$A$_2$ ground and excited states are split by 1.44 eV, the energy gap exceeds way above the thermal broadening for the working temperatures and is not relevant to the present discussion. On the other hand, the excited states $^4$A$_2$ and $^4$E which are split by only 4.4 meV corresponding to 51K should mix in a fashion that changes the optical emission properties. We see similar effects in the experiment as shown in Fig. S2 and below we give a simple quantum mechanical model that explains this temperature behavior.

Instead of considering all the 32 states of the excited subspace and ground state subspace, we consider a simpler case of a four-level system that explains the observed temperature behavior of the V1 excitation. The four-level system is formed by two excited states and two ground states. The Hamiltonian governing the dynamics in the dressed basis given by

$$H = E_+ |g_1\rangle\langle e_+| + E_- |g_2\rangle\langle e_-| + \lambda(|e_+\rangle\langle e_+| - |e_-\rangle\langle e_-|) + h.c.,$$

where the optical fields with an intensity $E_+$, couple the ground and excited states and $\lambda$, is the spin-orbit splitting of the excited states. The dressed excited states are given by $|e_\pm\rangle = \frac{1}{\sqrt{2}}[|e_1\rangle \pm |e_2\rangle]$.

Additionally, the excited states decay to the respective ground states and also suffer from a temperature dependent dephasing that causes incoherent mixing within the dressed basis. Including these non-unitary terms, the total dynamics is then by a master equation

$$\dot{\rho}(t) = -i[H, \rho(t)] + \sum_k \gamma_k \left[ \mathcal{L}_k \rho \mathcal{L}_k^\dagger - \frac{1}{2}\left(\rho \mathcal{L}_k^\dagger \mathcal{L}_k + \mathcal{L}_k^\dagger \mathcal{L}_k \rho\right)\right]$$

where the Lindblad operators $\mathcal{L}_1 = |g_1\rangle\langle e_+|$, $\mathcal{L}_2 = |g_2\rangle\langle e_-|$, $\mathcal{L}_3 = (|e_1\rangle\langle e_1| - |e_2\rangle\langle e_2|)$, with respective decay rates $\gamma_1$, $\gamma_2$, $\gamma_d$. While the optical decay is described by the operators $\mathcal{L}_1$, $\mathcal{L}_2$ the dephasing is described by $\mathcal{L}_3$. We assume a power-law temperature dependence for the dephasing rate, i.e., $\gamma_d(T^\alpha)$.

By taking $\gamma_2 \rangle \gamma_1$, and solving the master equation, we find the temperature dependence of the relative intensity of the two emission lines by evaluating the steady state populations in the ground states that are initially equally populated.

At very high temperatures, i.e., $\gamma_d \gg \lambda$, the dressed basis is completely dephased resulting in no population exchange among the two excited states. Similarly, in the other extreme limit $\gamma_d \ll \lambda$, there is no population exchange between $|e_+\rangle$ and $|e_-\rangle$ states as the detuning is much larger than the thermal dephasing. Only in the intermediate regime, there is population exchange between $|e_+\rangle$, allowing for maximal population transfer to $|g_2\rangle$, i.e., enhanced emission in the second decay channel, V1'.

Moreover, the physical origin of the mixing between the $^4A_2$ and $^4E$ states in our model can be determined by examining the molecular vibrations of the Si monovacancy center. These Raman and infrared active vibrational normal modes in the vicinity of the defect belong to the $2a_1 + 2e$ irreducible representations of the C$_{3v}$ symmetry group. Therefore, the first (V1) and second (V1') excited states with $^4A_2$ and $^4E$ symmetry, respectively, are only allowed to couple each other by the E symmetry vibrational normal modes. Such inter-level coupling between nearly degenerate electronic states that are linear in lattice displacements is also known as the pseudo-Jahn-Teller (PJT) effect. Using an electron-lattice coupling Hamiltonian of

$$H_{e-p} = G \begin{pmatrix} 0 & \theta & -\eta \\ \theta & 0 & 0 \\ -\eta & 0 & 0 \end{pmatrix} + \frac{\Delta}{3} \begin{pmatrix} -2 & 0 & 0 \\ 0 & 1 & 0 \\ 0 & 0 & 1 \end{pmatrix} + \frac{1}{2} K \left( \theta^2 + \eta^2 \right)$$

given in the electronic basis states $\{|A\rangle; |E_x\rangle, |E_y\rangle\}$ of $^4A_2$ and $^4E$ states, we obtain a minimum Jahn-Teller energy configuration of

$$E_{JT} = -\frac{\Delta}{6} - \varepsilon_0 - \left(\frac{\Delta}{4\varepsilon_0}\right)^2$$

at normal mode displacement $Q = \sqrt{\theta^2 + \eta^2} = \left[ Q_0^2 - (\Delta/2G)^2 \right]^{1/2}$. The $\varepsilon_0 = G^2/2K$ and $Q_0 = G/K$ correspond to the PJT energy and normal mode nuclear displacement, respectively, when the energy splitting between the excited states is zero ($\Delta = 0$). They are given in terms of the electron-phonon coupling $G$ and lattice $K$ force constants. From $Q > 0$, the condition for the PJT instability to occur reduces to the well-known order-of-magnitude rule $|\Delta/(4\varepsilon_0)| < 1$ [2]. It is important to note that due to small energy splitting (4.4 meV) between the excited states of this defect, this condition gets satisfied and the wave function at $A_2$ excited state minimum becomes a linear combination of the electronic $A_2$ and $E$ symmetry states.

To illustrate the temperature dependence of this PJT mixing, we define the coupling strength of the $E$ symmetry normal mode vibrations to the defect as $g(\omega) = \rho_D(\omega) f(q(\omega)) \xi(\omega)$ in terms Debye density of states $\rho_D = 3\omega^2/2\upsilon_n^3\pi^2$ and phonon coupling coefficient $\xi(\omega) = \left( \hbar \omega_n / 2M\upsilon_n^2 \right)^{1/2}$ for the $n^{th}$ mode with velocity $\upsilon_n$. This leads to a $T^{3/2}(\omega^{3/2})$ temperature dependence consistent with our findings above. The defect wave functions extended over many lattice sites are reflected in a cut-off function $f(q) = \left( 1 + r_B^2 q^2/4 \right)^{-2}$ using an effective mass approximation within an effective Bohr radius of $r_B$. We now have a quick look at one of the *E* symmetry transverse phonon modes around the defect propagating along [100] and polarized along [001] with sound velocity of $v = 7.1 \times 10^3$ m/s in 4H-SiC. We use an effective Bohr radius of 2.7 [3] nm to include most of the charge density around the V1 defect, comparable to some deep center acceptor states and 3C-SiC. Resulting electron-phonon coupling strength square with respect to temperature is shown in Fig. S3 and it is closely related to what is observed in the experiment (Fig. S2) since the contrast between V1 and V1' states is proportional to $g^2$. Although more sophisticated cut-off functions can be used for this deep center, they are beyond the scope of the current work.

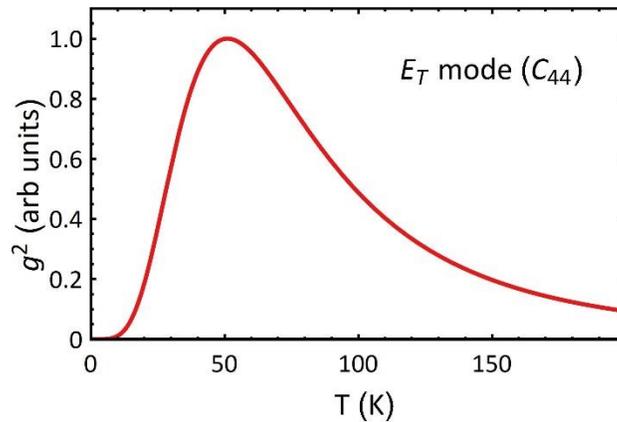

FIG. S3. Coupling strength of the E symmetry. Normalized electron-phonon coupling square $g^2$ between the E symmetry transverse vibrational mode (propagating along [001]) and $V_{Si}^-$ defect in 4H-SiC as a function of temperature.

## 3. Excited state lifetime

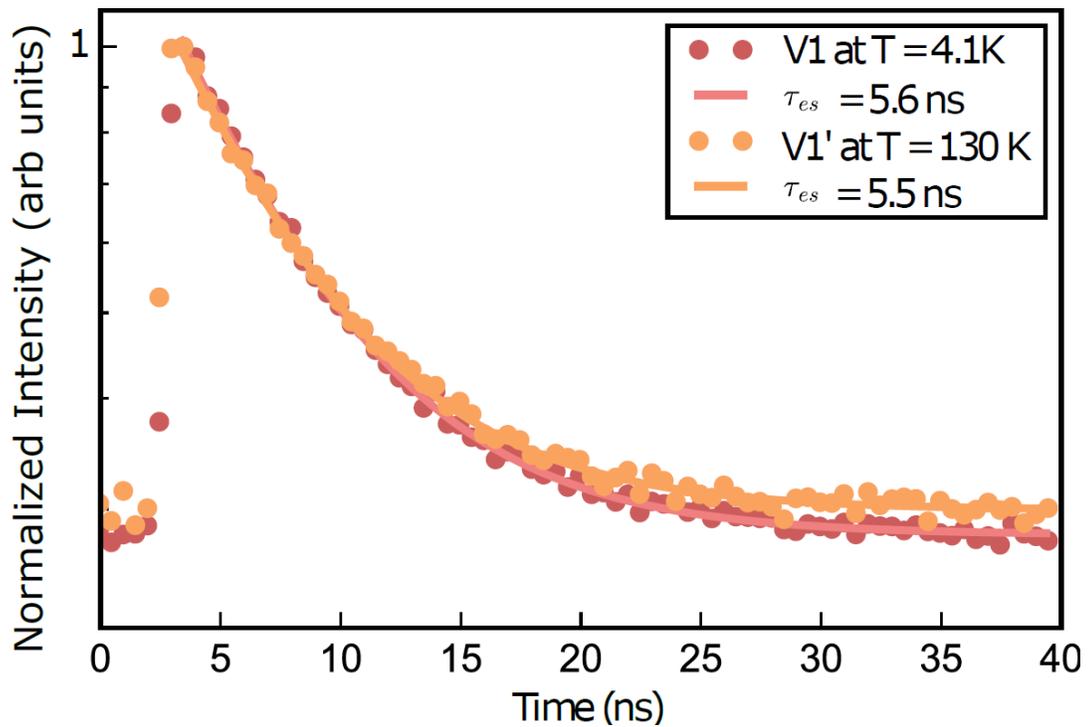

FIG. S4. The optically excited state lifetime is measured with an 805 nm picosecond laser.

In order to estimate the natural linewidth of the V1 and V1' transitions, the lifetimes of both excited states were measured by applying short laser pulses at 805 nm. To distinguish between the optical emission of V1 and V1' lines, we measured the excited state lifetime at different temperatures (5.5 K for V1, and 70 K for V1'). The observed decay curves are illustrated in Fig. S4. The excited state lifetimes of V1 and V1' are 5.5±1.4 ns and 5.6±1.2 ns, respectively. The lifetime of V1 is in agreement with the previously reported value [4] which determines the lower bound of the natural linewidth to be around 30 MHz.

## 4. Additional polarization dependence data

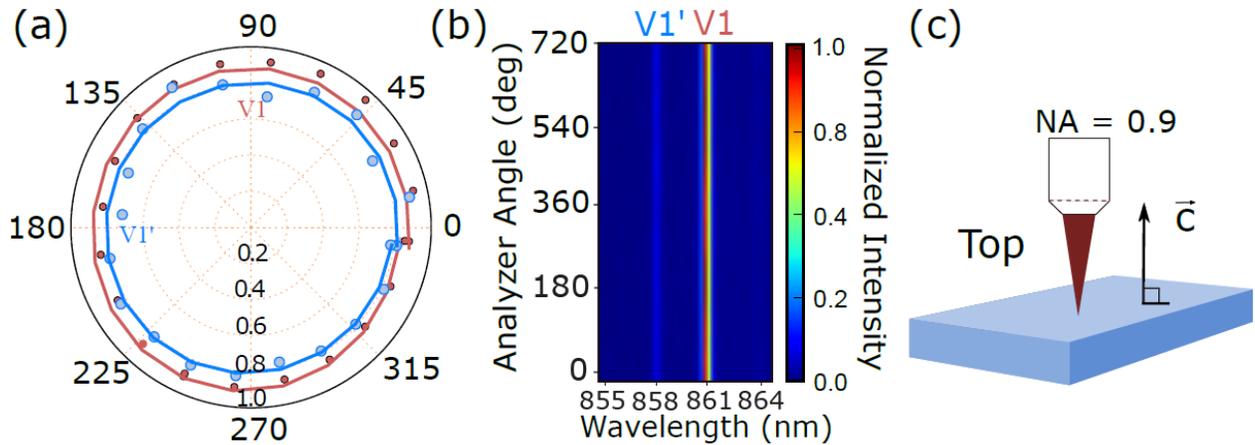

FIG. S5. The polarization of the V1/V1' ZPLs analyzed with an HWP and PBS at the sample orientation in which the laser incident orientation coincide with the c-axis. (a) The polar plot of the normalized V1 and V1' intensities. (b) The density plots showing the absolute intensities of the V1/V1' ZPLs. (c) The schematic diagrams depicting the sample orientation with respect to the laser incident orientation.

Since the sample has the c-axis perpendicular to its surface, the analysis of both $E \parallel c$ and $E \perp c$ polarizations of the emitted photons requires that the luminescence is excited and detected through the side surface of the sample (see the sketch in Fig. S5(c)). Figure S5 shows that only photons with $E \perp c$ polarization are registered in this geometry, and all directions in the plane perpendicular to the c-axis are equivalent, as illustrated by the polar plot. The polarization dependence data in FIG. S5 is almost independent of the rotation angle of the HWP. It indicates either an unpolarised light or a circularly polarized light. To identify the polarization status of the V1/V1' ZPL measured when the optical axis (laser incident orientation) was parallel to the c-axis, the additional analysis was done replacing the half-wave plate with a quarter-wave plate (QWP) before the PBS. As can be seen in Fig. S6, we observe the unpolarised light emission from both V1 and V1' ZPL.

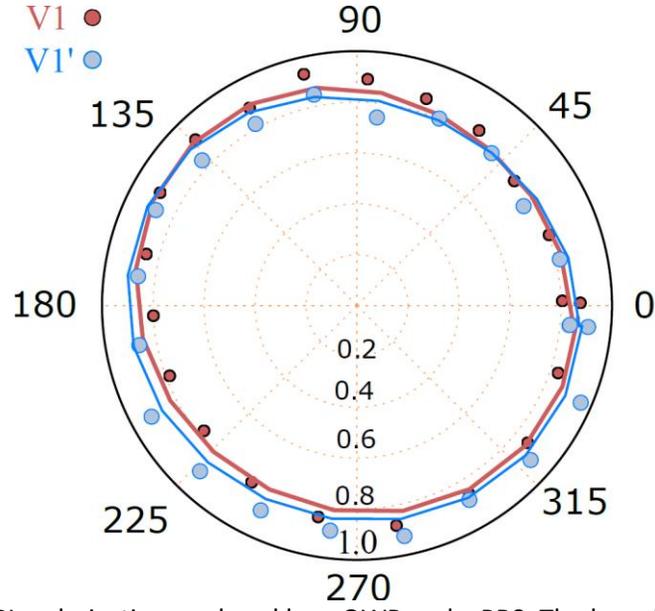

FIG. S6. V1 and V1' ZPL polarization analyzed by a QWP and a PBS. The laser incident orientation is parallel to the c-axis.

### 5. Group theory analysis of the V1 and V1' ZPL emission

In this section, we provide a detailed analysis of the observed polarization dependence of the V1 and V1' ZPLs shown in Fig. 1(c) of the main text and Fig. S5 and S6.

The individual symmetries of each state belonging to the V1 ground and first excited states (both labeled as $^4A_2$ in the manuscript due to their orbital symmetry) are given as $\{^1E_{3/2}, {}^2E_{3/2}, E_{1/2}^+, E_{1/2}^-\}$ in the $\{|3/2\rangle + i|-3/2\rangle, |3/2\rangle - i|-3/2\rangle, |1/2\rangle, |-1/2\rangle\}$ spin basis at zero magnetic field. Similarly, the symmetries of each state belonging to the second excited state (labeled as $^4E$) are given as $\{E_{1/2}^+, E_{1/2}^-, {}^1E_{3/2} + {}^2E_{3/2}, {}^1E_{3/2} - {}^2E_{3/2}, E_{1/2}^+, E_{1/2}^-, E_{1/2}^+, E_{1/2}^-\}$ in the $\{|3/2\rangle, |-3/2\rangle, |1/2\rangle, |-1/2\rangle, |1/2\rangle, |-1/2\rangle, |-3/2\rangle, |3/2\rangle\}$ spin basis. Note that the previously predicted small spin mixing between some of the $^4E$ states (due to higher order spin-spin interactions) [5] are omitted here. The optical selection rules between these various symmetries are given in Table S1 in terms of the vector components $\{(\hat{x},\hat{y});\hat{z}\} \in \{E;A_1\}$ of the electric field within the $C_{3v}$ double group. The $\hat{z}$ axis is parallel to the defect's c-axis.

| $C_{3v}$ | $E_{1/2}$ | $^1E_{3/2}$ | $^2E_{3/2}$ |
|---|---|---|---|
| $E_{1/2}$ | $A_1, E$ | $E$ | $E$ |
| $^1E_{3/2}$ | $E$ | - | $A_1$ |
| $^2E_{3/2}$ | $E$ | $A_1$ | - |

Table S1. Optical selection rules for the symmetries of $C_{3v}$ double group.

Therefore, by considering all four possible transitions after the spin selection rules are applied, we find that the V1 transition has a mixed polarization of $(E_\parallel : E_\perp)$ =(3:1) involving the electric field components parallel and orthogonal to the c-axis. For the more complicated V1' transition consisting of 12 possible transitions, we found the polarization ratio to be roughly (1:11). Thus, the V1 ($^4A_2$ to $^4A_2$) transition contains polarizations both parallel and orthogonal to the c-axis whereas the V1' ($^4A_2$ to $^4E$) is mostly polarized along the basal plane of the defect. These expectations are in very good agreement with our experimental results shown in Fig. 1(c) of the main text. Fig. S5 and S6 show that, as expected, only photons with E⊥c polarization are registered in this geometry, and all directions in the plane perpendicular to the c-axis are equivalent.

The orbital and spin properties of this defect are different from the well-studied NV center in diamond and the divacancy in SiC. While both ground states of V$_{Si}$ and NV centers feature $A_2$ orbital symmetry, V$_{Si}$ has entirely different electron and spin configuration (quartet versus triplet) with distinct optical selection rules dictated by the $\overline{C_{3V}}$ double group representations. The V1' transition with $E \perp c$ polarization is reminiscent of the NV center $E_{\pm} \rightarrow A_2$ transitions, which are circularly polarized and have been used for spin-photon entanglement [6].

## 6. Optical spin detection by V1' ZPL excitation

In this section, we present additional data about optically detected magnetic resonance (ODMR) experiments performed by resonant optical excitation of the V1' ZPL. For magnetic field alignment, we use the V2 centers in the ensemble, as described in refs. [7–9].

To resonantly excite V1' we used a home-built tunable external-cavity laser (λ = 858 nm) in continuous wave mode. Luminescence was detected with the 887 nm tunable long pass (LP) filter. The resultant relative ODMR intensity, calculated according to the explanation in the main text, is only ≈0.2 % and the sign is negative (Fig. S7). The change in the ODMR intensity by the resonant optical excitation of the V1' ZPL is significantly smaller than what we observed for the resonantly excited V1 line and comparable to the case of the off-resonant optical excitation (Fig. 2(b) in the main text). These observations could suggest that optical spin polarization is mainly established by the intersystem crossing (ISC) between the $^4A_2$ excited state, related to the optical line V1, and the $^2E$ metastable state. The ISC from the $^4E$ excited state, related to V1', does not seem to induce efficient optical polarization.

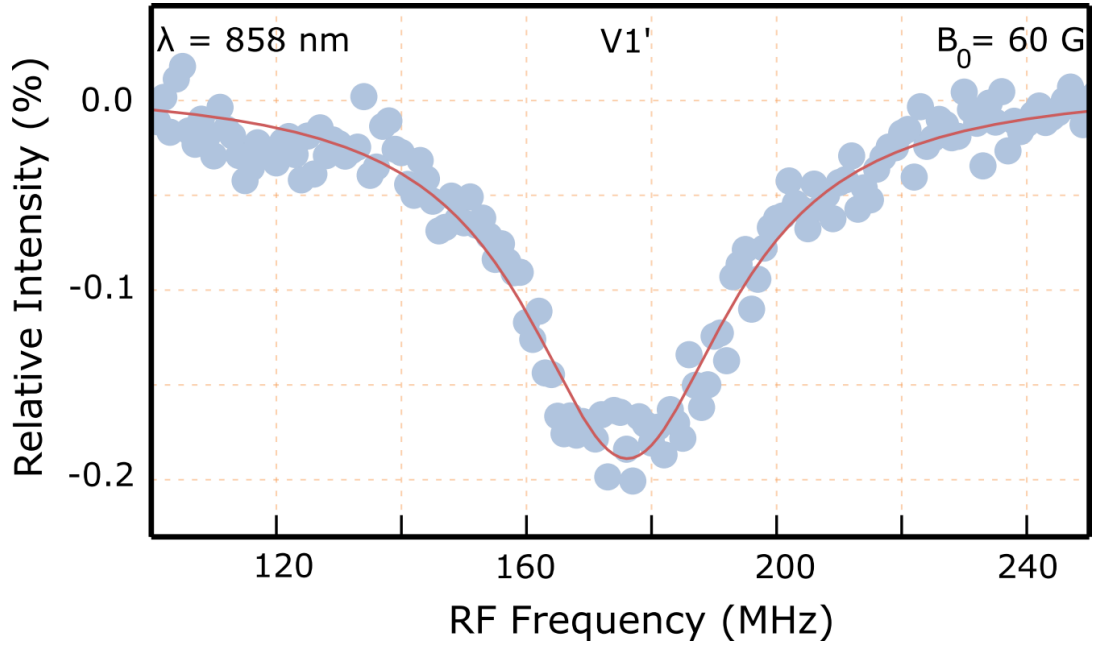

FIG. S7: ODMR under the resonant optical excitation of the V1' ZPL.

**7. Signal-to-Noise ratio of enhancement and quenching ODMR signal**

The contrast of ODMR signals is defined by $C \equiv (I_{max} - I_{min})/I_{max}$: $C_p = (I_p - I_0)/I_p$ and $C_m = (I_0 - I_m)/I_0$ for the signal which is enhanced and quenched by spin resonance, respectively. Here, $I_0$, $I_p$, and $I_m$ are the PL intensity without applying magnetic resonance, the total PL intensity which is enhanced and quenched by magnetic resonance, respectively. Note that $0 < C < 1$, and $C$ is different from the relative ODMR intensity used in Fig. 2 of the main text and Fig. S7. If we assume identical contrast ($C_p = C_m = C$), and shot-noise limit, the relative ratio of the signal-to-noise ratio (SNR) of the enhancement signal with respect to that of the quenching signal is, by using uncertainty propagation,

$$\frac{C_p/\delta C_p}{C_m/\delta C_m} = \frac{\delta C_m}{\delta C_p} = \frac{\sqrt{\frac{I_m I_0 + I_m^2}{I_0^3}}}{\sqrt{\frac{I_0 I_p + I_0^2}{I_p^3}}} = \frac{1}{\sqrt{1-C_m}} > 1 \ .$$

Thus, the enhancement signal has a better SNR ratio by a factor of $(1-C)^{-0.5}$

## 8. Spin Rabi oscillations

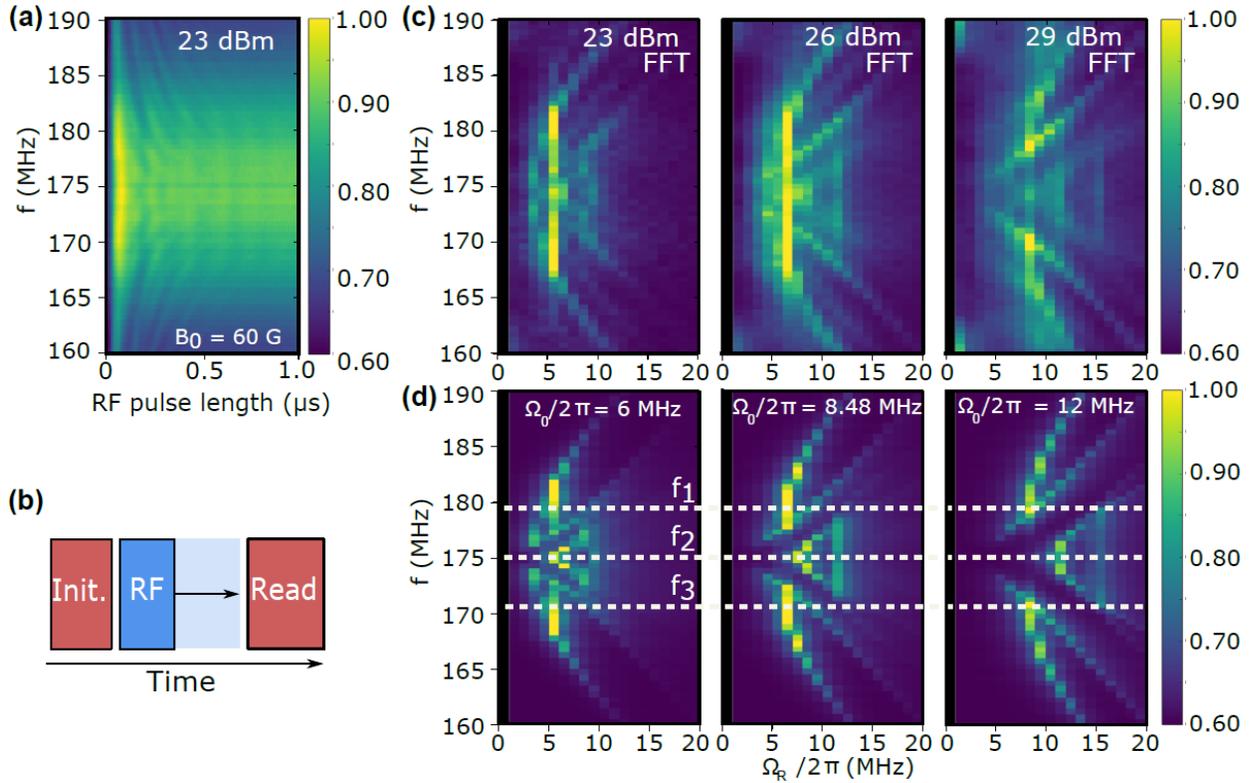

FIG. S8. (a) Rabi measurement with detuned RF driving frequencies. (b) Pulse scheme for a Rabi measurement. The first laser pulse (Init.) is polarizing the spin state. The RF pulse is manipulating the spin state followed by the last laser pulse (Read) for the spin state readout. (c) Fast Fourier transformed Rabi oscillations at different RF powers. (d) Simulated Rabi oscillations. The dotted lines indicate three resonant RF frequencies shown in Fig. 2(a). The strong zero frequency intensities in both (C) and (D) are removed for better distinguishability of the Rabi frequencies.

Here we discuss a theoretical model for the observed distribution of the Rabi oscillations as in FIG. 3 of the main text. For the readers' convenience, the main text figure 3 is duplicated here as FIG. S8.

**System Hamiltonian.** We consider a spin $S=3/2$ system, driven by a radio-frequency field with amplitude $\Omega$ and frequency $\omega$ under a static magnetic field $B$. The system dynamics is investigated assuming initial polarization into an incoherent mixture of $|S_z = \pm 3/2\rangle$. The system Hamiltonian is $H = H_0 + \Omega\left[\cos(\omega t)\hat{S}_x + \sin(\omega t)\hat{S}_y\right]$, where $H_0$ is the diagonal Hamiltonian with energy levels $\varepsilon_{\pm 3/2} = \pm\frac{3}{2}\gamma B$ and $\varepsilon_{\pm 1/2} = (D \pm \gamma B)$, $\gamma$ is the gyromagnetic ratio ($\gamma = 28$ MHz/mT) and $D$ the zero-field splitting ($2D = 4$ MHz for the V1 center).

We set ourselves in a frame rotating at angular velocity $\omega$ around the z-axis:

$$U = e^{-i\omega t \hat{S}_z} = \begin{bmatrix} e^{3i\omega t/2} & 0 & 0 & 0 \\ 0 & e^{i\omega t/2} & 0 & 0 \\ 0 & 0 & e^{-i\omega t/2} & 0 \\ 0 & 0 & 0 & e^{-3i\omega t/2} \end{bmatrix}$$

The Hamiltonian in this rotating frame can be calculated as $H_{rot} = U^\dagger H U - i\dot{U}U^\dagger$. Neglecting fast components at $2\omega$ (rotating-wave approximation), we get:

$$H_{rot} = \begin{bmatrix} \varepsilon_{-3/2} + (3/2)\omega & (\sqrt{3}/4)\Omega & 0 & 0 \\ (\sqrt{3}/4)\Omega^* & \varepsilon_{-1/2} + (1/2)\omega & \Omega/2 & 0 \\ 0 & \Omega^*/2 & \varepsilon_{+1/2} - (1/2)\omega & (\sqrt{3}/4)\Omega \\ 0 & 0 & (\sqrt{3}/4)\Omega^* & \varepsilon_{+3/2} - (3/2)\omega \end{bmatrix}$$

**System evolution.** Let $|u_i\rangle$ be the eigenvectors of $H_0$, i.e. the eigenvectors corresponding to $S_z$=-3/2, -1/2, +1/2, +3/2. The time-independent Schrödinger equation for $H$ is $H|v_i\rangle = \alpha_i |v_i\rangle$. Let $V$ the matrix of the eigenvectors of: $|v_i\rangle = \sum_k V_{ik}|u_k\rangle$.

Let us assume that the initial state is one of the eigenvectors of $H_0$: $|\psi_0\rangle = |u_k\rangle$, which can be expressed as a linear combination of eigenvectors of $H$ as $|u_k\rangle = \sum_l V_{kl}^{-1} |v_l\rangle$.

The temporal evolution is described by Schrödinger equation as:

$$|\psi(t)\rangle = e^{-iHt/\hbar} |\psi_0\rangle = \sum_l e^{-i\alpha_l t/\hbar} V_{kl}^{-1} |v_l\rangle = \sum_{l,m} e^{-i\alpha_l t/\hbar} V_{kl}^{-1} V_{lm} |u_m\rangle.$$

After a time $t$, the probability associated with the state $|u_m\rangle$ is:

$$\rho_{mk}(t) = \left| \sum_{l,m} e^{-i\alpha_l t/\hbar} V_{kl}^{-1} V_{lm} \right|^2.$$

Assuming that each spin level $|u_m\rangle$ is associated with a photoluminescence intensity $I_m$, the total photo-luminescence $I_k(t)$, assuming initialization into $|\psi_0\rangle = |u_k\rangle$, evolves as:

$$I_k(t) = \sum_m I_m \rho_{mk}(t).$$

For the silicon vacancy, optical excitation results in polarization into an incoherent mixture of $S_z = -3/2$ and $S_z = +3/2$. This case would require the solution of the Liouville equation for the density matrix. Here, we take a more empirical approach, considering the incoherent superposition of pure states, each evolving according to Schrödinger equation, as discussed above. Assuming that the probability of initial spin polarization into the eigenstate $|u_k\rangle$ is described by $P_k$, the temporal evolution of photo-luminescence intensity is calculated as:

$$I(t) = \sum_{m,k} I_m P_k \left| \sum_{l,m} e^{-i\alpha_l t/\hbar} V_{kl}^{-1} V_{lm} \right|^2$$

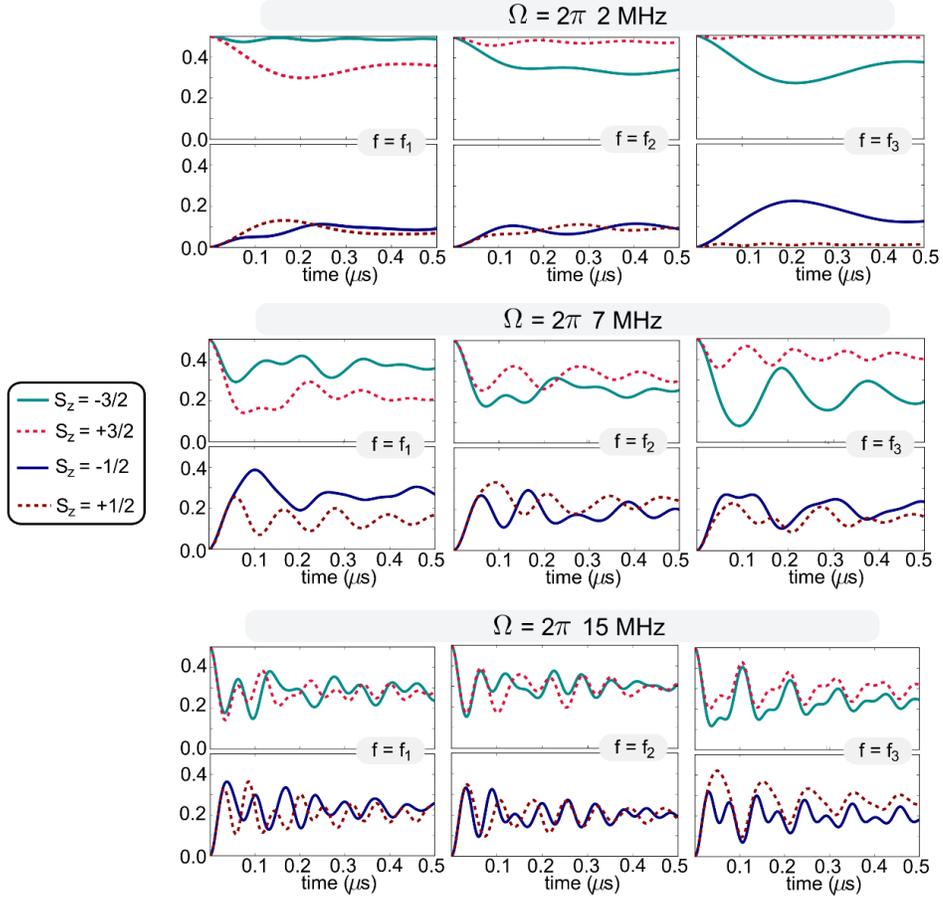

FIG. S9. Rabi model. The probability of occupation for the four spin levels as a function of time under radiofrequency driving at frequency $f$ and amplitude $\Omega$.

**Numerical simulations**. We investigate the electron spin dynamics assuming perfect initial polarization in an incoherent mixture of Sz=-3/2 and $S_z$=+3/2, corresponding to $P_{+3/2} = P_{-3/2} = 0.5$, $P_{+1/2} = P_{-1/2} = 0$. Results are shown in Fig. S9, for three different values of $\Omega$ ($\Omega/2\pi$=2MHz, 7MHz, and 15MHz). In each case, we plot the dynamics for a driving frequency corresponding to each of the three transitions: $f = f_1$ ($|-3/2\rangle \leftrightarrow |-1/2\rangle$), $f = f_2$ ($|-1/2\rangle \leftrightarrow |+1/2\rangle$), $f = f_3$ ($|+1/2\rangle \leftrightarrow |+3/2\rangle$). We include decoherence in the results by an exponential decay with time constant $T_2^*$ =200 ns.

The simulation results, presented in Fig. 3(d) of the main text and Fig. S8(d), show good agreement with the experimental data. Simulation accuracy could be further improved by including inhomogeneous broadening and finite bandwidth of the applied RF field. Numerical simulations were also performed solving the Liouville-von Neumann equation for the density matrix: no significant difference was observed with respect to the simplified model illustrated above.

### 9. Pulsed spin control

Due to the incoherent mixture of various spin states during the Rabi driving, we find that at various driving strengths, there are no perfect revivals of the population in any subspace of the four-level system. This makes the definition of π, and π/2- pulses more complex when compared to the well-studied two-level system. For both experimental and theoretical analysis, we define a π-pulse as the minimum time required for transferring the maximal population from $|\pm 3/2\rangle \rightarrow |\pm 1/2\rangle$ subspace. Experimentally, this is signaled by the maximum contrast in the observed fluorescence. But to really know how the state of the four-level system looks during these pulses, we consider the following limits

(i) $\Omega \gg D$, i.e., when the applied $B_1$ field strength is much stronger than the zero-field splitting. In this limit, the π-pulse will result in

$$|\pm 3/2\rangle \rightarrow \frac{1}{\sqrt{2}}\left(|\pm 3/2\rangle \pm i |\mp 1/2\rangle\right)$$

and similarly the π/2-pulse,

$$\left|\pm 3/2\right\rangle \rightarrow \frac{1}{\sqrt{3}}\left(\sqrt{2}\left|\pm 3/2\right\rangle \mp i\left|\pm 1/2\right\rangle\right)$$

(ii) $\Omega \approx D$, i.e., when the applied $B_1$ field strength is of the same order as that of the zero-field splitting, the π-pulse will result in

$$\left|\pm 3/2\right\rangle \rightarrow \frac{1}{3}\left[\left(\sqrt{2}\left|\pm 3/2\right\rangle - \left|\mp 3/2\right\rangle\right) + \sqrt{3}\left(\left|\pm 1/2\right\rangle \pm i\left|\mp 1/2\right\rangle\right)\right]$$

(iii) $\Omega \ll D$, i.e., when the applied microwave power is much smaller than that of the zero-field splitting, the π-pulse will result in

$$\left|\pm 3/2\right\rangle \rightarrow i\left|\pm 1/2\right\rangle$$

The final states after the applied pulses clearly depict the complexity involved incoherently driving and controlling a four-level system.

### 10. Spin Rabi oscillations at high magnetic field

The Rabi measurements were done at $B_0 = 60$ and $1000$ G. The data which were collected at $B_0 = 60$ G is shown in Fig. 2 (c,d,e) of the main text and Fig. S8. The magnetic field was applied at $B_0 = 60$ G with a large cylindrical permanent magnet (30 mm diameter and 50 mm thick) from outside the cryostat chamber, and the magnetic field was aligned to the c-axis.

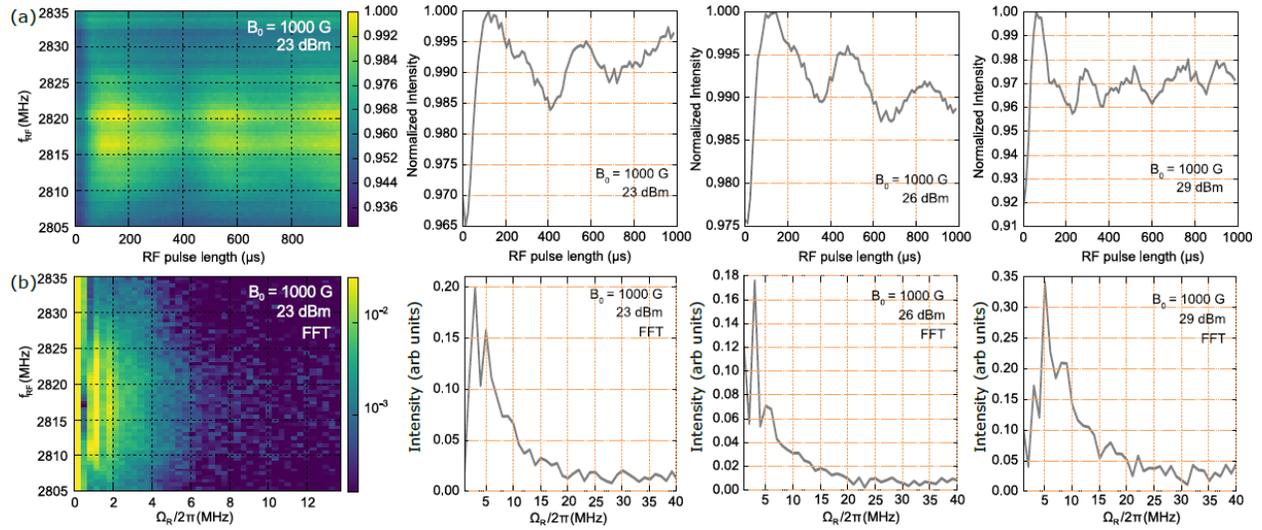

FIG. S10. Rabi oscillations at $B_0$= 1000 G. The density plots: The observed Rabi oscillations at RF power of 23 dBm as a function of the RF frequency (a) and the FFT of them (b). The 2D plots: The Rabi oscillations at the RF frequency of 2820 MHz measured at three different RF powers (a) and their FFT (b).

For $B_0$=1000 G a small samarium cobalt magnet (1 mm thick and 10 mm diameter) was placed beneath our 4H-SiC sample. The measured Rabi oscillations in the time domain and their FFT can be seen in Fig. S10. Seriously broaden Rabi frequency distributions near-independent on the detuning of the RF field frequency over the tested RF frequency range can be seen. We attribute this broadening to the serious inhomogeneous broadening originated from the small magnet placed right below the sample. Note that similar broadening can happen in the disordered spin distribution under a homogeneous magnetic field [10,11]. The large magnet used for 60 G experiment may not produce such serious inhomogeneity since it was placed outside the chamber (17 cm away from the sample) thus the magnetic field gradient inside the focal volume could not be significant.

## 11. Spin decoherence and dephasing measurements

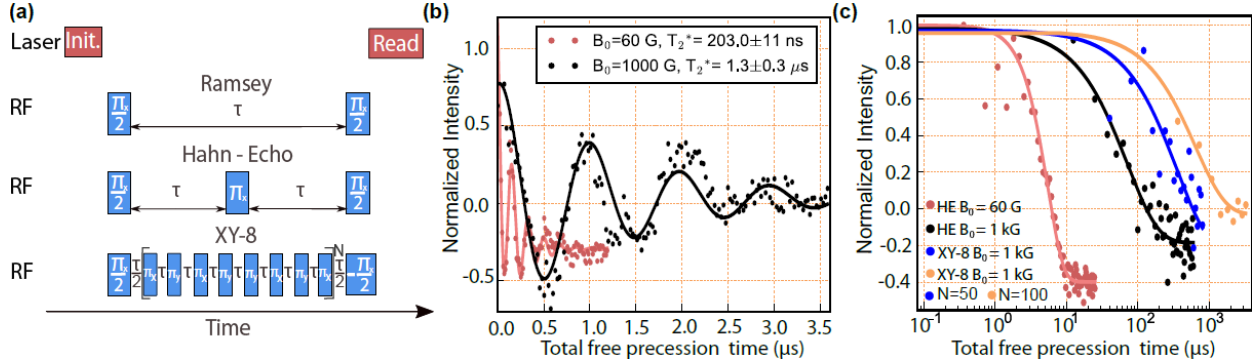

FIG. S11. (a) Ramsey measurement at two different magnetic fields $B_0$= 60 and 1000 G. (b) The spin decoherence measured at two magnetic fields $B_0$= 60 and 1000 G. The measured spin decoherence time $T_2$ is 4.4 μs and 83 μs at 60 and 100 G, respectively. The spin decoherence times by the XY-8 decoupling sequence are 286.5 μs and 0.6 ms with $N$=10 and 50, respectively.

In the main text, we presented and discussed the spin coherence investigated only at $B_0$ $B_0$=1000 G. We indeed studied spin coherence at T= 5.5 K with both $B_0 = 60$ G and 1000 G by Ramsey, Hahn-echo, and XY-8 dynamical decoupling pulse sequences. Spin measurements were performed by optical excitation resonant with the V1 ZPL (861 nm). As discussed above, coherent spin control poses a challenge for a four-level system with a small ZFS. In the following experiments, we take the duration of a π-pulse as the minimum time required for transferring the maximal population from $|\pm 3/2\rangle$ to $|\pm 1/2\rangle$ subspace. Experimentally, this is signaled by the maximum contrast in the observed Rabi oscillation signals.

The Ramsey pulse sequence can be seen in Fig. S11(a). After optical spin polarization, we applied a π/2 pulse to create a superposition between $|\pm 3/2\rangle$ and $|\pm 1/2\rangle$ sublevels. Another π/2 pulse was applied after a free precession time τ for a projective readout before the readout laser pulse. We observed an evolution of the coherent superposition, with a frequency corresponding to the detuning from the resonant transition, and spin dephasing time $T_2^* = 200 \pm 11$ ns and $1.3 \pm 0.3$ μs at $B_0$ =60 G and 1000 G, respectively, as shown in Fig. S11(b). To suppress the inhomogeneous broadening in an ensemble and decouple the spin ensembles from low frequency spin noise sources, we applied a Hahn-Echo sequence. The Hahn-Echo pulse sequence can be seen in Fig. S11(a), which adds a π pulse between two π/2 pulses, to refocus the dephased spin ensemble due to inhomogeneous broadening and slowly fluctuating magnetic noise. Although the applied π pulses exhibit limited spin control to a single transition as discussed above, we could see a typical exponential decay with decoherence time $T_2 = 4.4 \pm 0.3$ μs at $B_0$ =60 G (Fig. S11(c)). Since a drastically improved coherence time is expected at a stronger $B_0$ field by suppressing nuclear spin flip-flops due to a large mismatch of nuclear spin Zeeman levels of $^{29}$Si and $^{13}$C [12], we applied a high magnetic field along the c-axis (0001). The spin decoherence time increases at $B_0$ =1000 G to $T_2 = 83.9 \pm 1.6$ μs. The observed $T_2$ is, however, shorter than the theoretical expectations [12] and the value measured for a single V2 center at room temperature [7]. This could be related to the imperfect π pulses and the inhomogeneity of the $B_0$ field for the case $B_0$ = 1000 G as explained above. These observations, however, do support the findings by Carter *et al.* [13], related to the fact that the initial state, dephased during the free-precession, cannot be refocused by a π pulse due to the oscillating local fields produced by coupled nuclear spins. Thus, the shorter $T_2$ could be related to electron spin echo envelope modulation (ESEEM) [14]. The four sublevels of a S=3/2 electronic spin have four different non-zero values of the hyperfine coupling to nearby nuclear spins and thus result in more complex ESEEM than S=1 systems, whose sublevels have only two different non-zero coupling values [7,12]. Furthermore, as reported by Carter *et al.*, the ensemble inhomogeneous broadening induces beating oscillations among the various modulation frequencies, leading to a shortening of measured by Hahn-echo [13]. To further suppress decoherence, we applied the XY-8 dynamical decoupling sequence, which acts as a filter for the environmental magnetic noise [15]. This sequence has proven to be effective to extend the coherence time of the S=3/2 spin ensemble associated with the V2 center from the nuclear spin bath in 4H-SiC [16]. A repeated decoupling pulse scheme leads to a better suppression of noise, increasing the spin decoherence time with N=10 and N=50 repetitions to a value of respectively $T_2 = 286 \pm 7$ μs and $T_2 = 0.60 \pm 0.01$ ms (Fig. S11(c)).

## 12. Single V1 silicon vacancy center.

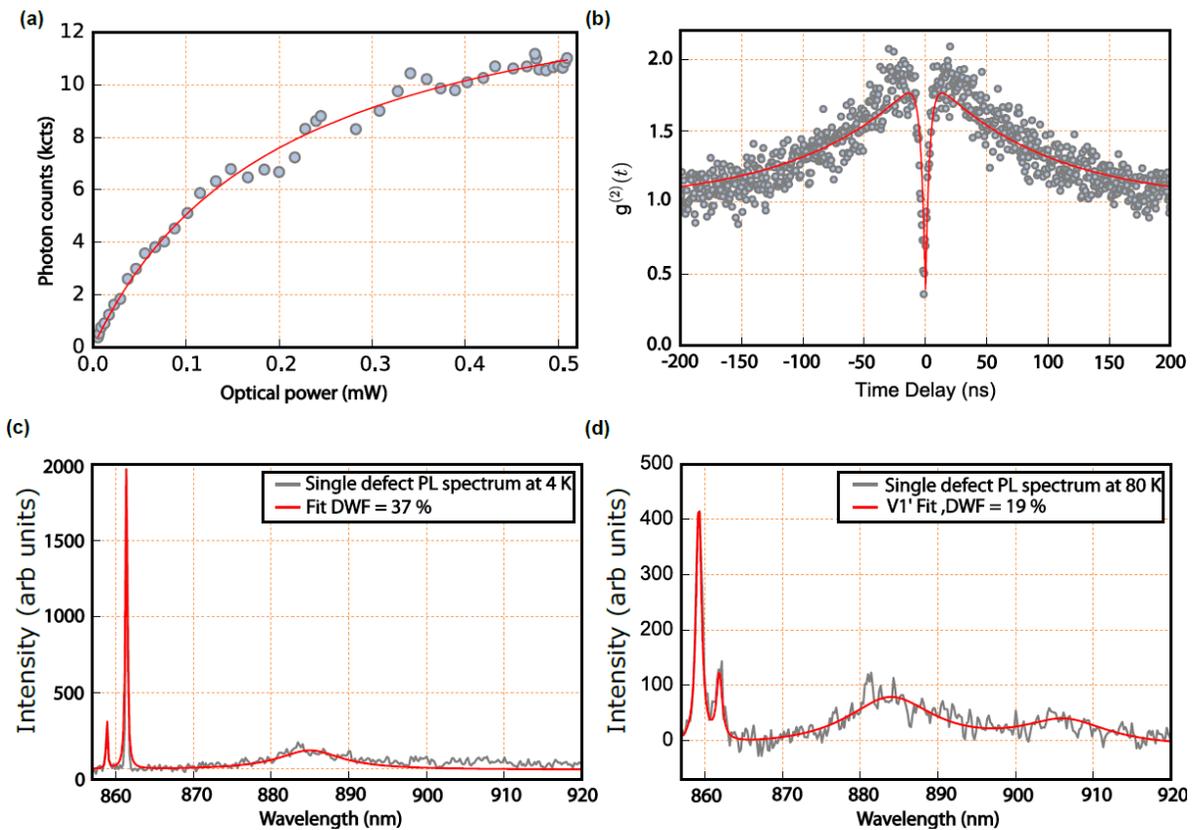

FIG. S12. (a) The optical saturation curve of a single V1 center PL emission. (b) The $g^{(2)}$ autocorrelation measurement indicates clearly a single photon emission character, $g^{(2)} < 0.5$. The data in (a) and (b) were taken from the single V1 center shown in the main text Fig.4. (c, d) The PL spectra of a single V1 center at 4 K and 80 K, respectively. Note that this V1 center is different from one in the main text Fig. 4.

We started our investigation for the single V1 center with a pillar sample [17]. The pillars act as a broadband optical waveguide, which provides enhanced photon collection efficiency. The 4H-SiC pillar sample was electron irradiated (2 MeV) with an electron dose of $4\times10^{14}$ cm$^{-2}$. We found statistically a single V1 or V2 defect in one out of 20 pillars. The density of single V1 defects is higher compared to single V2 defects. The ratio between single V1 and V2 center is roughly 3:1. The saturation power of a single defect inside a pillar is 190 µW, and the saturated count rate is 14 kcps. The photon detection efficiency in the tested wavelength range (around 900 nm) is roughly a half of the optimized wavelength (around 700 nm) of this Si-APD based detector. Since the saturated count rate of a single V1 center measured at 5 K with an air objective was 14 kcps, one can expect ~30 kcps if we assume the optimized efficiency at the tested wavelength.

The Debye-Waller factor (DWF) of the V1 zero phonon line was measured at 4 K where the phonon side band caused by V1' is expected to be minimum in our experimental conditions. The formula how the DWF for V1 is calculated is $\frac{\int V1}{\int V1 + \int PSB}$. The best measured DWF of V1 is 40% as in Fig. 4 of the main text. The lowest DWF measured from another single V1 center is 37% as shown in Fig. S12 (c). At 80 K, at which V1 is minimized, the measured DWF for V1' is 19 % (Fig. S12(d)). There is uncertainty in our data because we don't know how strong the PSB caused by V1' and V1 is at 4 K and 80 K, respectively. Both are assumed to be zero at 4 K and 80 K, respectively. This indicates that the real DWF of V1 and V1' can be higher. If we calculate the DWF at 4 K as the overall number of coherent photons in comparison to the incoherent photons $\frac{\int (V1+V1')}{\int (V1+V1') + \int PSB}$ we achieve a DWF of 48 % and 41 %, using the data in Fig. 4 of the main text and the data at 4K in Fig. S12, respectively. The DWF at 80K calculated with this formula is 32%.

*sangyun.lee@kist.re.kr

†These authors contributed equally to this work.